\documentclass[aps,prd,reprint,a4paper,showpacs,nofootinbib,superscriptaddress,floatfix]{revtex4-2}

\usepackage{amsmath,amssymb,bm}
\usepackage{mathtools}

\usepackage{graphicx}
\usepackage[utf8]{inputenc}
\usepackage[T1]{fontenc}
\usepackage{txfonts}
\usepackage{mathrsfs}
\usepackage{xcolor}
\usepackage{booktabs}
\usepackage{placeins}
\usepackage{hyperref}
\hypersetup{
  colorlinks=true,
  linkcolor=red,
  citecolor=blue,
  urlcolor=blue,
  pdftitle={CMB Test of the Higgs Origin of Dark-Photon Dark Matter},
  pdfauthor={Imtiaz Khan, Salvatore Capozziello, G. Mustafa, Chengxun Yuan, and Farruh Atamurotov}
}
\newcommand{\Ap}{A'}
\newcommand{\Pcal}{\mathcal P}
\newcommand{\be}{\beta_{\rm iso}}
\newcommand{\as}{A_s}
\newcommand{\order}{\mathcal O}
\newcommand{\qeff}{q_{\rm eff}}
\newcommand{\hh}{\bm h}
\newcommand{\hbarv}{\bar{\bm h}}
\newcommand{\vev}{v_h}

\begin{document}
\title{CMB Test of the Higgs Origin of Dark-Photon Dark Matter}

\author{Imtiaz Khan}
\email{ikhanphys1993@gmail.com}
\affiliation{Department of Physics, Zhejiang Normal University, Jinhua, Zhejiang 321004, China}
\affiliation{Research Center of Astrophysics and Cosmology, Khazar University, Baku, AZ1096, 41 Mehseti Street, Azerbaijan}

\author{Salvatore Capozziello}
\email{capozziello@na.infn.it}
\affiliation{Dipartimento di Fisica ``E. Pancini", Universit\`a di Napoli ``Federico II", Complesso Universitario di Monte Sant’ Angelo, Edificio G, Via Cinthia, I-80126, Napoli, Italy,}
\affiliation{Istituto Nazionale di Fisica Nucleare (INFN), sez. di Napoli, Via Cinthia 9, I-80126 Napoli, Italy,}
\affiliation{Scuola Superiore Meridionale, Largo S. Marcellino, I-80138, Napoli, Italy.}

\author{G. Mustafa}
\email{gmustafa3828@gmail.com}
\affiliation{Department of Physics, Zhejiang Normal University, Jinhua, Zhejiang 321004, China}

\author{Chengxun Yuan}
   \email{yuancx@hit.edu.cn}
\affiliation{School of Physics, Harbin Institute of Technology, Harbin 150001, People's Republic of China}

\author{Farruh~Atamurotov}
\email{atamurotov@yahoo.com}
\affiliation{School of Physics, Harbin Institute of Technology, Harbin 150001, People's Republic of China}
\affiliation{University of Tashkent for Applied Sciences, Str. Gavhar 1, Tashkent 100149, Uzbekistan}
\affiliation{National University of Uzbekistan, Tashkent 100174, Uzbekistan}

\begin{abstract}
Laboratory searches determine the mass and kinetic mixing of dark-photon dark matter. The CMB isocurvature probes how its abundance depends on inflationary initial conditions. We formulate this dependence through the logarithmic response $q_{\rm eff}=\partial\ln n_{A'}/\partial\ln r_i$ of the final vector number to the primordial dark-Higgs displacement. For a smooth local abundance map, the separate-universe relation connects $q_{\rm eff}$ to the CDM isocurvature. The uncorrelated Planck limit applies when the dark Higgs is a light and energetically subdominant spectator with weak inflaton mixing. In the quadratic conserved-number branch displacement-independent transfer gives $q_{\rm eff}=2$. The all-dark-matter bound then requires $r_i/H_*>3.5\times10^4$. We calculate the dynamical map by evolving the radial mode and its tangent response through a smooth Abelian-Higgs crossover. The post-transition orbit gives the comoving action. The map contains positive, negative, and near-zero responses at finite transferred action. Tanh and error-function crossovers retain these branches. A thermal-mass profile and finite inflationary quartics shift their locations continuously. Stochastic duration and radiative stability then determine realization regions. Early symmetry breaking and momentum redshift add the remaining conditions. Low-energy measurements and the CMB response can distinguish Higgsed-vector cosmologies with identical present-day $(m_{A'},\epsilon)$ and relic abundance.

\end{abstract}

\maketitle

\section{Introduction}

A future dark-photon signal could determine a mass and kinetic mixing while leaving the relic origin open. A Higgsed vector carries an additional cosmological coordinate through the sensitivity of its final abundance to the inflationary displacement of the field that generates its mass. CMB isocurvature probes this differential dependence and thereby tests the dark-Higgs transfer history \cite{Khan:2025ibo}.

Among the wide class of dark matter candidates, dark photons have attracted considerable attention as well-motivated extensions of the Standard Model. In the conventional Stueckelberg realization, the dark photon is characterized at low energies by two effective parameters, namely its mass and kinetic mixing with the visible photon. Such scenarios naturally accommodate several non-thermal production mechanisms, including inflationary production and vacuum misalignment, which can successfully generate a relic population of vector dark matter \cite{Nelson:2011sf,Arias:2012az,Graham:2015rva}. 

A qualitatively different situation emerges when the dark photon mass originates from a Higgs mechanism associated with the spontaneous breaking of a hidden $U(1)$ gauge symmetry. In this Higgsed framework, the physical description necessarily involves additional scalar-sector dynamics, including the structure of the symmetry-breaking potential, the cosmological history of the hidden-sector phase transition, and the radiative stability of the scalar potential itself \cite{Holdom:1985ag,Ahlers:2008qc,An:2013yua,Caputo:2021eaa,Redi:2022zkt}. As a consequence, the final dark-photon abundance depends on low-energy portal parameters together with the detailed cosmological history governing the hidden scalar sector. Inflationary vector production, axion-assisted amplification, resonant particle generation, quenched energy transfer, cosmic-string production, and coherent vector evolution therefore define distinct cosmological histories capable of generating equal relic abundance \cite{Agrawal:2018vin,Co:2018lka,Dror:2018pdh,Long:2019lwl,Nakayama:2019rhg,Kitajima:2023xhn,Kitajima:2024vbc}.

Cosmological isocurvature provides a probe of this production history. It is well known from the classical axion misalignment mechanism that when a light scalar field determines the final dark matter abundance, quantum fluctuations generated during inflation induce entropy perturbations that survive as observable isocurvature signatures in the cosmic microwave background (CMB) \cite{Preskill:1982cy,Abbott:1982af,Dine:1982ah,Linde:1985yf,Seckel:1985tj,Lyth:1989pb}. Similar theoretical arguments have subsequently been extended to a broad class of spectator-field, freeze-in, and non-thermal hidden-sector dark matter scenarios \cite{Starobinsky:1994bd,Hardwick:2017fjo,Markkanen:2018bfx,Garcia:2023awt,Bellomo:2022pav,Holst:2023lrm}. 

The physical mechanism follows from the separate evolution of superhorizon patches. If the final dark matter abundance depends on the amplitude of a light scalar field displaced during inflation, then neighboring superhorizon patches acquire slightly different dark matter densities due to inflationary fluctuations of that scalar field. The relevant observable is therefore the logarithmic derivative of the final abundance with respect to the primordial displacement. This derivative links the inflationary scalar fluctuation to the subsequent transfer history \cite{Pirzada:2026jml,Pirzada:2026sle,Pirzada:2026uak}.

Higgsed dark-photon models realize this mechanism. Previous studies analyzed inflationary fluctuations and abundance transfer in the dark-Higgs sector. They also examined isocurvature, backreaction, symmetry restoration, and defect formation \cite{Redi:2022zkt,Cline:2024mdg,Kitajima:2024vbc,Cyncynates:2023defectfree,Cyncynates:2024targets}. We calculate the abundance-response map from the primordial radial displacement to the final vector number. Power-law limits provide reference branches. The continuous crossover determines the response between these limits.

We calculate this map for a perturbative Abelian-Higgs crossover. The radial background and tangent system evolve together. The post-transition orbit defines a conserved comoving action and the final vector number after field-independent perturbative decay. The two-dimensional response surface contains positive, negative, and near-zero $q_{\rm eff}$ at finite transferred action. Tanh and error-function crossovers retain all three branches. A thermal mass proportional to $a^{-2}$ and finite inflationary quartics shift their locations continuously. The value $q_{\rm eff}=2$ is recovered in the quadratic conserved-inheritance limit. Combining the calculated response with CMB isocurvature tests cosmological histories that share $(m_{A'},\epsilon)$ and the present abundance. The response definition also applies to non-equilibrium resonance maps \cite{Khan:2026higgsedresonance}.

The separate-universe relation follows from a smooth single-valued abundance map on a uniform-radiation hypersurface. Its uncorrelated CMB spectrum additionally requires a light and subdominant spectator with weak inflaton--dark-Higgs mixing. The inequality $q_{\rm eff}\ge2$ applies to quadratic oscillations with conserved comoving scalar number and a nondecreasing transfer factor. The crossover profile determines the transfer derivative. Inflationary priors and particle-cosmology conditions then specify a realization of the calculated response.

Section~\ref{sec:framework} derives the separate-universe relation and states the inflationary spectator conditions. Section~\ref{sec:inheritance} defines the quadratic conserved-inheritance branch and its transfer criterion. Section~\ref{sec:dynamical} calculates the response generated by the Abelian-Higgs crossover. Section~\ref{sec:applications} applies the stochastic prior and scalar-sector conditions. It then examines daughter momentum, early symmetry breaking, and portal constraints.

\section{Response framework}
\label{sec:framework}
\subsection{Inflationary Higgs phase}

During the inflationary epoch, the scalar dynamics governing the hidden sector can differ from the late-time symmetry-breaking structure responsible for generating the dark-photon mass. In particular, the stochastic evolution of the dark Higgs field can be controlled by an effective inflationary potential that differs substantially from the conventional Mexican-hat potential realized after reheating. To describe this framework, we consider the minimal Higgsed dark-photon sector governed by the Lagrangian

\begin{equation}
\begin{aligned}
\mathcal L\supset& -\frac14F'_{\mu\nu}F'^{\mu\nu}-\frac14B_{\mu\nu}B^{\mu\nu}
+\frac{\epsilon_Y}{2}F'_{\mu\nu}B^{\mu\nu}+|D_\mu\Phi|^2\\
&-V_{\rm inf}(|\Phi|)-V_{\rm br}(|\Phi|),
\end{aligned}
\label{eq:modelL}
\end{equation}

where $A'_\mu$ denotes the dark gauge boson associated with the hidden $U(1)$ symmetry and $B_{\mu\nu}$ is the Standard Model hypercharge field strength. Above electroweak symmetry breaking the interaction $F'_{\mu\nu}B^{\mu\nu}$ is the gauge-invariant renormalizable portal between the two Abelian factors. After electroweak symmetry breaking $B_{\mu\nu}=c_WF_{\mu\nu}-s_WZ_{\mu\nu}$ and
\begin{equation}
\frac{\epsilon_Y}{2}F'_{\mu\nu}B^{\mu\nu}
=\frac{\epsilon_Yc_W}{2}F'_{\mu\nu}F^{\mu\nu}
-\frac{\epsilon_Ys_W}{2}F'_{\mu\nu}Z^{\mu\nu},
\qquad \epsilon\equiv\epsilon_Yc_W.
\label{eq:ewmatching}
\end{equation}
For $|\epsilon_Y|\ll1$ and $m_{A'}\ll m_Z$ the coefficient $\epsilon$ is the photon-mixing parameter used in low-energy constraints. Canonical normalization gives relative corrections of order $\epsilon_Y^2$. The induced neutral-current coupling is suppressed by $m_{A'}^2/m_Z^2$ \cite{Holdom:1985ag,Caputo:2021eaa}. The homogeneous cosmological background has vanishing Standard Model gauge field strengths. The radial background and tangent equations therefore retain their form at this order. The gauge-covariant derivative is $D_\mu=\partial_\mu+i g_D A'_\mu$. The complex scalar responsible for spontaneous symmetry breaking is parameterized as $\Phi=(h_1+i h_2)/\sqrt2$ with $\hh=(h_1,h_2)$ and $r=|\hh|$. The hidden-sector gauge coupling is $g_D\equiv q_H e_H$.

During inflation, the radial scalar degree of freedom evolves under an effective local potential which, in the vicinity of the relevant field configuration, can be written as

\begin{equation}
V_{\rm inf}(r)=V_0+\frac12m_+^2 r^2+\frac14\lambda_+ r^4+\cdots,
\qquad m_+^2\ge0,
\label{eq:Vinf}
\end{equation}

where the effective mass parameter $m_+^2$ generally receives contributions from curvature effects, Hubble-induced corrections, or other ultraviolet-scale interactions present during inflation. At later cosmological times, the hidden symmetry undergoes spontaneous breaking through a low-energy potential of the form

\begin{equation}
V_{\rm br}(r)=\frac{\lambda_h}{4}(r^2-\vev^2)^2+\cdots,
\qquad m_{\Ap}=g_D\vev .
\label{eq:Vbr}
\end{equation}

The $U(1)$-invariant abundance depends on the radial magnitude. Decomposing the inflationary perturbation into radial and tangential components gives $\delta r=\hat{\bm n}\!\cdot\!\delta\hh$ at linear order, while the angular component rotates the background direction and first enters $\delta r$ at quadratic order. The radial projection therefore contains the linear abundance isocurvature source. In the Higgs phase the angular degree of freedom becomes the longitudinal vector polarization.

The response formalism uses a smooth single-valued map from the primordial radial displacement to the final vector abundance. This map permits a local separate-universe comparison between neighboring Hubble patches. It can include adiabatic crossover, thermal release, entropy evolution, and non-equilibrium transfer through the corresponding $\rho_{A'}(r)$ map. The central numerical surface uses the quadratic limit of Eq.~\eqref{eq:Vinf}. Appendix~\ref{app:robustness} treats finite $\lambda_+$ and a thermal-mass transition.

\subsection{General separate-universe response}

To quantify the cosmological imprint of inflationary dark-Higgs fluctuations, we consider the scalar field decomposition $\Phi=(h_1+i h_2)/\sqrt2$, with field-space vector $\hh=(h_1,h_2)$ and radial amplitude $r=|\hh|$. For a generic production history, the final dark-photon energy density evaluated on a uniform-radiation hypersurface is assumed to depend on the inflationary scalar displacement through the functional relation $\rho_{\Ap}(r)$. We therefore define the logarithmic abundance-response parameter as

\begin{equation}
\qeff(r)\equiv \frac{\partial\ln\rho_{\Ap}}{\partial\ln r} .
\label{eq:qeff}
\end{equation}

This quantity measures the retained primordial information in the final dark-photon abundance. Once the stable vector population is nonrelativistic, $\rho_{\Ap}=m_{\Ap}n_{\Ap}$ and this response equals $\partial\ln(a^3n_{\Ap})/\partial\ln r$ for constant $m_{\Ap}$. For a local Hubble patch satisfying $\hbarv\ne0$, the separate-universe formalism implies that the resulting cold-dark-matter entropy perturbation can be expressed as

\begin{equation}
S_c=3(\zeta_c-\zeta_\gamma)
=f_{\Ap}\qeff \frac{\hat{\bm n}\cdot\delta\hh}{|\hbarv|}
+\order\!\left(\frac{\delta h^2}{|\hbarv|^2}\right),
\qquad
\hat{\bm n}=\frac{\hbarv}{|\hbarv|},
\label{eq:linearS}
\end{equation}

assuming that all remaining dark matter components evolve adiabatically. Here $\zeta_c$ and $\zeta_\gamma$ denote the curvature perturbations associated with cold dark matter and radiation, respectively, $\delta\hh$ represents the inflationary fluctuation of the dark Higgs field, and $f_{\Ap}\equiv\Omega_{\Ap}/\Omega_c$ gives the fractional contribution of dark photons to the total cold dark matter abundance.

The numerical analysis adopts the separable inflationary limit $\delta V=0$ in
$V_{\rm tot}(\varphi,r)=V_\varphi(\varphi)+V_{\rm inf}(r)+\delta V(\varphi,r)$. A weakly coupled inflationary completion remains in the spectator regime when
\begin{equation}
\begin{aligned}
\frac{|m_{r,{\rm eff}}^2|}{H_*^2}&\ll1,
&\Omega_D^{\rm inf}&\equiv\frac{\rho_h+\rho_{A'}}{3M_{\rm Pl}^2H_*^2}\ll1,\\
\frac{|\partial_\varphi\delta V|}{|V_\varphi'|}&\ll1,
&\frac{|V_{,\varphi r}|}{H_*^2}&\ll1,
&\frac{\omega_\perp}{H_*}&\ll1.
\end{aligned}
\label{eq:spectatorconditions}
\end{equation}
Here $m_{r,{\rm eff}}^2\equiv V_{{\rm tot},rr}$ and $\omega_\perp\equiv|D_t\hat{\bm\sigma}|$ is the bending rate of the background field trajectory. In the separable limit $m_{r,{\rm eff}}^2=m_+^2+3\lambda_+|\hbarv|^2$. At leading slow-roll order the light-mass condition gives $\Pcal_{\delta r}(k_*)=(H_*/2\pi)^2$. The hidden-sector energy fraction controls the correction to the Friedmann equation. The inflaton-force ratio controls the correction to the inflaton equation. The mixed Hessian and bending rate suppress mode mixing at horizon crossing and conversion outside the horizon. Any inflaton--dark-Higgs interaction is contained in $\delta V$.

Under these conditions $\Pcal_{\zeta\,\delta r}\simeq0$ \cite{Hardwick:2017fjo,Markkanen:2018bfx}. Equation~\eqref{eq:linearS} then gives
\begin{equation}
\Pcal_{S_c}(k_*)=f_{\Ap}^2\qeff^2
\left(\frac{H_*}{2\pi|\hbarv|}\right)^2 .
\label{eq:master}
\end{equation}
For finite inflaton--dark-Higgs mixing,
\begin{align}
\Pcal_{\zeta S_c}&=\frac{f_{\Ap}q_{\rm eff}}{|\hbarv|}\,
\Pcal_{\zeta\,\delta r},\\
\cos\Delta&\equiv
\frac{\Pcal_{\zeta S_c}}{\sqrt{\Pcal_\zeta\Pcal_{S_c}}}.
\label{eq:crossspectrum}
\end{align}
The numerical bound in Eq.~\eqref{eq:Planckbound} uses $\cos\Delta=0$. A finite value of $\cos\Delta$ is tested with the correlated CDM-isocurvature likelihood. The abundance derivative in Eq.~\eqref{eq:qeff} remains defined by the background abundance map. The CMB likelihood then carries the correlation dependence \cite{Akrami:2018odb}.

This structure parallels axion misalignment, where $\rho_a\propto\theta_i^2$ gives a logarithmic response of two. For a Higgsed vector, the fluctuating radial displacement also governs the symmetry-breaking and transfer history. A quartic condensate entering a quadratic oscillatory phase gives the analytic inheritance value $q_{\rm eff}=3$. The numerical transition map below evaluates the quadratic central branch and its finite-quartic deformations.

Using the standard isocurvature fraction parameter defined as $\be=\Pcal_{S_c}/(\as+\Pcal_{S_c})$, together with the measured scalar amplitude $\as=2.10\times10^{-9}$ and the Planck 2018 upper bound on uncorrelated CDM isocurvature $\be<0.038$ at 95\% confidence \cite{Akrami:2018odb,Aghanim:2018eyx}, one obtains the constraint

\begin{align}
\frac{|\hbarv|}{H_*}>1.75\times10^4\,|\qeff| f_{\Ap},\qquad
|\qeff| f_{\Ap}\frac{\sqrt{\Pcal_h}}{|\hbarv|}<9.11\times10^{-6}.
\label{eq:Planckbound}
\end{align}

The resulting response parameter space is illustrated in the left panel of Fig.~\ref{fig:entropy}. For the benchmark case in which dark photons constitute the full dark matter abundance and quadratic conserved inheritance gives $|\qeff| f_{\Ap}=2$, the uncorrelated Planck limit requires a minimum local scalar displacement $|\hbarv|/H_*>3.5\times10^4$. Representative future sensitivities $\beta_{\rm iso}=10^{-2}$ and $3\times10^{-3}$ correspond to displacement scales $6.9\times10^4$ and $1.27\times10^5$, respectively \cite{Dvorkin:2022bsc,Montandon:2020axv,Valiviita:2017fbx}.

The CMB constraint probes the relative inflationary fluctuation of the abundance-generating scalar through $q_{\rm eff}H_*/|\hbarv|$. The matched photon-mixing parameter $\epsilon$ enters laboratory observables separately. The measured scalar amplitude and the abundance derivative set the local consistency condition. A statistical prior enters only when a distribution for the local displacement is specified.

\section{Conserved perturbative inheritance}
\label{sec:inheritance}
The quadratic conserved-number branch defines the analytic reference for the full response calculation. At the onset of oscillations,
\begin{equation}
\rho_h=\frac12m_h^2r_i^2,\qquad
N_h\equiv a^3n_h=\frac{a^3\rho_h}{m_h}\propto r_i^2.
\label{eq:initialnumber}
\end{equation}
For a subdominant condensate quadratic evolution conserves $N_h$ until the decay epoch. On a uniform-radiation hypersurface the entropy density and the final vector mass are common to neighboring patches. The logarithmic response can therefore be evaluated from the vector yield. After decay the yield has the form
\begin{equation}
Y_{A'}(r_i)=C\,r_i^2\mathcal F(r_i),\qquad
q_{\rm eff}=2+\frac{\partial\ln\mathcal F}{\partial\ln r_i}.
\label{eq:qboundmain}
\end{equation}
The coefficient $C$ is independent of $r_i$. The factor $\mathcal F$ contains vector multiplicity, branching fractions, transferred fraction, and entropy dilution. In the reference branch these factors are independent of $r_i$ and give $q_{\rm eff}=2$. A nondecreasing transfer factor gives $q_{\rm eff}\ge2$ when $\partial\ln\mathcal F/\partial\ln r_i\ge0$. A decreasing transfer factor gives $q_{\rm eff}<2$. Saturation, dilution, and thermalization provide examples. Equation~\eqref{eq:qboundmain} defines the domain of the quadratic conserved-inheritance criterion.

As a normalization check we evolve
\begin{align}
\frac{d\rho_R}{dN}&=-4\rho_R,\nonumber\\
\frac{d\rho_h}{dN}&=-3\rho_h-\frac{\Gamma_h}{H}\rho_h,\nonumber\\
\frac{d\rho_{A'}}{dN}&=-3\rho_{A'}+\frac{\Gamma_h}{H}\rho_h,
\label{eq:boltzmain}
\end{align}
with $H^2=(\rho_R+\rho_h+\rho_{A'})/(3M_{\rm Pl}^2)$ and $\rho_h(N_i)=m_h^2r_i^2/2$. A symmetric finite difference in $r_i$ returns $q_{\rm eff}=2.00$ for the conserved-number reference branch.

\section{Dynamical abundance-response map}
\label{sec:dynamical}
The crossover calculation evaluates the transfer-function derivative beyond the monotone conserved-inheritance branch while retaining the separate-universe relation in Eq.~\eqref{eq:linearS}. Let $x=r/v_h$, measure energies in units of $m_h^2v_h^2$, and center the symmetry-breaking transition at $N=0$. The reference crossover is

\begin{align}
U(x,N)&=[1-s(N)]\left(\frac{\mu_+^2x^2}{2}+\frac{\Lambda_+x^4}{4}\right)
+s(N)\frac{(x^2-1)^2}{8},\nonumber\\
s(N)&=\frac12\left[1+\tanh\left(\frac{N}{\Delta N}\right)\right],
\label{eq:smoothtransition}
\end{align}
where $\mu_+=m_+/m_h$ and $\Lambda_+=\lambda_+v_h^2/m_h^2$.  The central map uses $\Lambda_+=0$, while Appendix~\ref{app:robustness} evaluates finite quartics, an error-function crossover, and a radiation-era thermal mass $m_T^2\propto T^2\propto a^{-2}$ \cite{Quiros:1999jp}.  During radiation domination, define $\mathcal H(N)=H(N)/m_h$, with $\mathcal H'/\mathcal H=-2$.  The background and tangent systems are
\begin{align}
x''+x'+\frac{U_{,x}}{\mathcal H^2}&=0,\label{eq:radialtransition}\\
u_x'&=u_z,\qquad
u_z'=-u_z-\frac{U_{,xx}}{\mathcal H^2}u_x,
\label{eq:tangentflow}
\end{align}
where $u_x=\partial x/\partial\ln r_i$ and $u_z=\partial x'/\partial\ln r_i$.

After the crossover, a controlled orbit in one broken-phase well has action
\begin{equation}
J(E)=\frac1\pi\int_{x_-}^{x_+}dx\,
\sqrt{2[E-U_{\rm br}(x)]},\qquad \mathcal I_h=a^3J(E).
\label{eq:actionmap}
\end{equation}
Once $H/\omega\ll1$, $\mathcal I_h$ is the conserved comoving adiabatic invariant.  Field-independent perturbative decay gives $a^3n_{A'}\propto\mathcal I_h$, and hence
\begin{equation}
q_{\rm eff}=\frac{J_{,E}}{J}
\left[\mathcal H^2x'u_z+U_{{\rm br},x}u_x\right].
\label{eq:qdynamic}
\end{equation}
This quantity is the logarithmic Jacobian of the transfer map. Positive values describe local stretching of nearby initial conditions. The line $q_{\rm eff}=0$ marks local focusing toward equal final comoving action. Negative values reverse the abundance ordering. Each branch occurs at finite transferred action.

The calculation includes smooth single-basin trajectories with $E<0.96E_b$. The action handoff is performed at $H/\omega<0.1$. This domain has $r_i/v_h=\mathcal O(1)$ and gives the CMB-implied symmetry-breaking scale point by point. Light vector masses remain compatible with this scale through $g_D=m_{A'}/v_h$. Mode-resolved transfer maps use this logarithmic observable for separatrix-crossing and resonance histories \cite{Redi:2022zkt,Cyncynates:2023defectfree,Pirzada:2026npl,Khan:2026nsz,Khan:2026higgsedresonance}.

\begin{figure*}[!t]
\centering
\includegraphics[width=\textwidth]{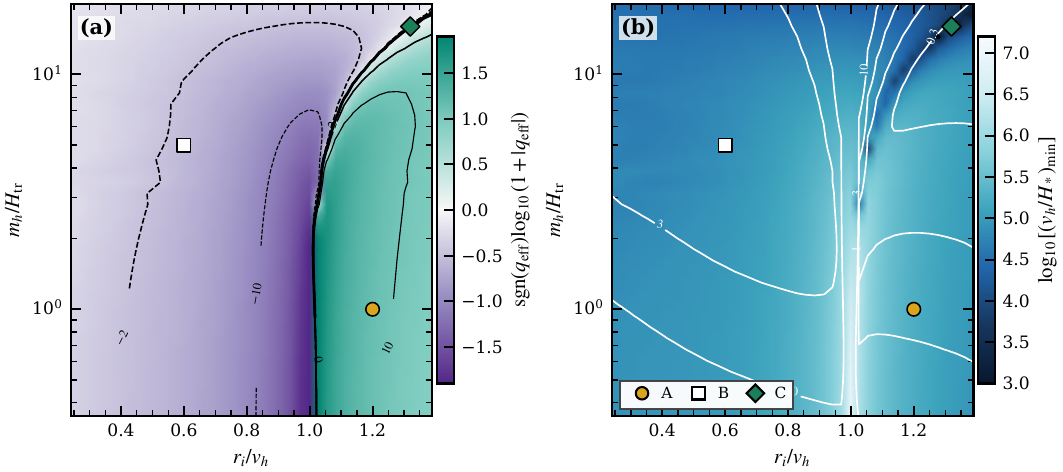}
\caption{Dynamical abundance response through the dark-Higgs crossover. Panel (a) gives the signed logarithmic response from the tangent system and post-transition action map. The central contour marks $q_{\rm eff}=0$. Panel (b) gives the minimum $v_h/H_*$ required by the Planck isocurvature limit for $f_{A'}=1$. White contours give the comoving action relative to the sudden-transition action. The black contour marks $|q_{\rm eff}|=0.1$. Markers identify the amplified (A), sign-reversed (B), and response-reduced (C) trajectories.}
\label{fig:responsemap}
\end{figure*}

For $f_{A'}=1$, the local CMB condition becomes
\begin{equation}
\left(\frac{v_h}{H_*}\right)_{\rm min}
=\frac{1.7475\times10^4|q_{\rm eff}|}{r_i/v_h}.
\label{eq:vhcmb}
\end{equation}
Figure~\ref{fig:responsemap} displays the calculated $q_{\rm eff}$ and the corresponding CMB lower bound for $\mu_+=0.2$ and $\Delta N=0.15$. The response surface contains preserved, amplified, sign-reversed, and response-reduced regions. The normalized-action contours show finite abundance transfer along the near-zero corridor. The markers correspond to an amplified trajectory $(q_{\rm eff}=12.486)$, a sign-reversed trajectory $(-2.583)$, and a response-reduced trajectory $(-0.096)$. Numerical convergence and tangent-response comparisons are given in Appendix~\ref{app:numerics}.

\begin{table*}[t]
\caption{Representative dynamical transition histories.  The normalized action is the final comoving action divided by the sudden-transition action at equal $r_i/v_h$ with $a_{\rm tr}=1$.  The final column follows from Eq.~\eqref{eq:vhcmb} for $f_{A'}=1$.}
\label{tab:dynamic}
\scriptsize
\begin{ruledtabular}
\begin{tabular}{lcccccc}
Branch & $r_i/v_h$ & $m_h/H_{\rm tr}$ & $q_{\rm eff}$ & normalized action & $H/\omega$ & $(v_h/H_*)_{\rm min}$ \\
\colrule
A amplified & $1.20$ & $1$ & $12.486$ & $2.108$ & $3.00\times10^{-2}$ & $1.82\times10^5$ \\
B sign reversed & $0.60$ & $5$ & $-2.583$ & $1.346$ & $1.82\times10^{-2}$ & $7.52\times10^4$ \\
C response reduced & $1.32$ & $16$ & $-0.096$ & $0.248$ & $5.67\times10^{-3}$ & $1.27\times10^3$ \\
\end{tabular}
\end{ruledtabular}
\end{table*}
\section{Cosmological realization conditions}
\label{sec:applications}
A cosmological realization specifies the inflationary prior and radial potential. It also specifies the symmetry-breaking epoch, decay kinematics, and visible portal. These inputs determine which region of the response map is realized. The local relation between $S_c$ and $q_{\rm eff}$ retains the form derived above. The numerical regions below use the stated benchmark assumptions.

\subsection{Perturbative benchmark and daughter momentum}
The daughter momentum gives an additional cosmological coordinate. For $r_h=m_h/m_{A'}\gg1$, two-body decay produces $p_d\simeq m_h/2$, followed by
\begin{equation}
\frac{p(T)}{m_{A'}}\simeq \frac{r_h}{2}\frac{T}{T_d}
\left[\frac{g_{*s}(T)}{g_{*s}(T_d)}\right]^{1/3} .
\label{eq:coldness}
\end{equation}
Abundance normalization, CMB response, and the nonrelativistic transition provide three consistency coordinates for a Higgs-origin relic \cite{Bode:2000gq,Viel:2013apy,Irsic:2017ixq,Villasenor:2022aiq}.

\begin{figure*}[t]
\centering
\includegraphics[width=\textwidth]{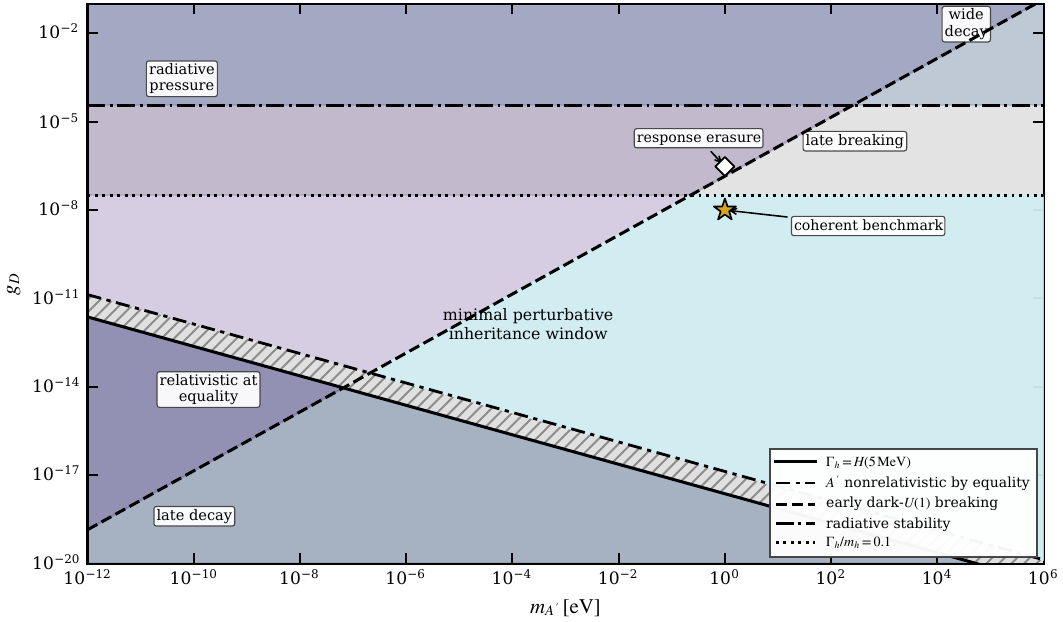}
\caption{Minimal perturbative Higgs-origin vector dark matter for \(r_h=m_h/m_{\Ap}=10^8\). The colored regions identify decay below \(5\,\mathrm{MeV}\) and relativistic evolution at equality. They also identify delayed dark-\(U(1)\) breaking, radiatively sensitive quartics, and broad-width decay. The blue region is the joint consistency window. The gold star is the coherent benchmark of Table~\ref{tab:bench}. The white diamond is a saturated-yield branch where the scalar response is reduced before freeze-out.}
\label{fig:central}
\end{figure*}

Figure~\ref{fig:central} shows the model-space projection for the minimal all-dark-matter benchmark $r_h=10^8$. The blue window satisfies decay before $5\,\mathrm{MeV}$ and nonrelativistic evolution by equality. It also satisfies early hidden-$U(1)$ breaking, radiative stability of the minimal quartic, and the narrow-width condition $\Gamma_h/m_h<0.1$. The hatched strip marks the momentum boundary between timely decay and cold behavior. Varying $r_h$ moves this strip with the decay boundary. The calculated $q_{\rm eff}$ determines the corresponding CMB scalar-sector boundary.

\begin{table*}[t]
\caption{Three representative outcomes. The table separates a coherent inherited displacement on the CMB boundary from a short stochastic draw with an exponentially small tail probability. It also gives a saturated yield with reduced scalar response. For the saturated branch $Y_{\rm inh}$ is the inherited yield. The quantities $Y_{\rm sat}$ and $\mathcal E$ denote the attractor yield and integrated erasure strength.}
\label{tab:bench}
\scriptsize
\begin{ruledtabular}
\begin{tabular}{llll}
Branch & Input & Response & Consequence \\
\colrule
Coherent perturbative & $m_{A'}=1\,{\rm eV}$, $g_D=10^{-8}$, $m_h/m_{A'}=10^8$ & $q_{\rm eff}=2.00$ & on the current Planck boundary \\
Short stochastic & quadratic branch, $N_{\rm st}=60$, typical $r/H_*=1.74$ & $q_{\rm eff}=2$ & $P_{\rm pass}=e^{-4.02\times10^8}$ \\
Saturated yield & $Y_{\rm inh}/Y_{\rm sat}=10$, $\mathcal E=12$ & $q_{\rm eff}=1.23\times10^{-4}$ & $h_{\rm iso}/H_*=2.15$, $P_{\rm pass}=0.22$ \\
\end{tabular}
\end{ruledtabular}
\end{table*}

Figure~\ref{fig:responsemap} calculates $q_{\rm eff}$ from the crossover dynamics. Figure~\ref{fig:central} projects the quadratic value $q_{\rm eff}=2$ onto the stated particle-cosmology benchmark. The first is a dynamical response map. The second is a realization of its quadratic branch.

\subsection{Stochastic prior and finite duration}

The local isocurvature constraint derived in Eq.~(\ref{eq:Planckbound}) is local and therefore precedes a statistical prior for specific inflationary histories such as finite-duration stochastic misalignment or spectator-field relaxation dynamics \cite{Graham:2018jyp,Chatrchyan:2022dpy}. In the case of quadratic equilibrium, each real scalar component of the dark Higgs acquires the well-known inflationary variance

\begin{equation}
\sigma_h^2=\frac{3H_*^4}{8\pi^2m_h^2},
\qquad
P(\hh)\propto\exp\!\left[-\frac{\hh^2}{2\sigma_h^2}\right].
\label{eq:quadvar}
\end{equation}

For a complex dark Higgs field, replacing the local displacement variable \(|\hbarv|^2\) by the typical equilibrium radial expectation value leads to the useful order-of-magnitude estimate

\begin{equation}
\Pcal_{S_c}^{\rm typ}=\frac13(\qeff f_{\Ap})^2\left(\frac{m_h}{H_*}\right)^2,
\qquad
\frac{m_h}{H_*}<\frac{1.58\times10^{-5}}{|\qeff| f_{\Ap}}.
\label{eq:massbound}
\end{equation}

For the quadratic conserved-inheritance all-dark-matter benchmark characterized by \(|\qeff| f_{\Ap}=2\), this condition becomes \(m_h/H_*<7.9\times10^{-6}\). The radial tail of the stochastic distribution then gives the probability for a patch to satisfy the local displacement criterion,

\begin{equation}
P_{\rm pass}=\exp\!\left[-\frac{h_{\rm iso}^2}{2\sigma_h^2}\right],
\qquad
h_{\rm iso}=1.75\times10^4|\qeff| f_{\Ap}H_* .
\label{eq:Ppass}
\end{equation}

For an inflationary stochastic stage of finite duration, the scalar variance evolves according to

\begin{equation}
\sigma_h^2(N_{\rm st})=\frac{3H_*^4}{8\pi^2m_h^2}
\left[1-\exp\!\left(-\frac{2m_h^2N_{\rm st}}{3H_*^2}\right)\right],
\label{eq:finite}
\end{equation}

which defines the relaxation timescale $N_{\rm rel}=3H_*^2/(2m_h^2)$.

At the quadratic conserved-inheritance all-dark-matter equilibrium boundary, the relaxation time is \(N_{\rm rel}\simeq2.4\times10^{10}\). In the finite-duration light-field regime, \(\sigma_h^2\simeq H_*^2N_{\rm st}/(4\pi^2)\), and \(N_{\rm st}=60\) gives
$P_{\rm pass}\simeq\exp(-4.0\times10^8)$ for \(|\qeff|f_{\Ap}=2\).
This finite-duration benchmark therefore requires a coherently prepared mean displacement. Since $h_{\rm iso}^2\propto\beta_{\rm iso}^{-1}$, a projected sensitivity \(\beta_{\rm iso}=10^{-2}\) gives \(P_{\rm pass}=\exp[-1.6\times10^9]\) for this duration.

\begin{figure*}[!t]
\centering
\includegraphics[width=\textwidth]{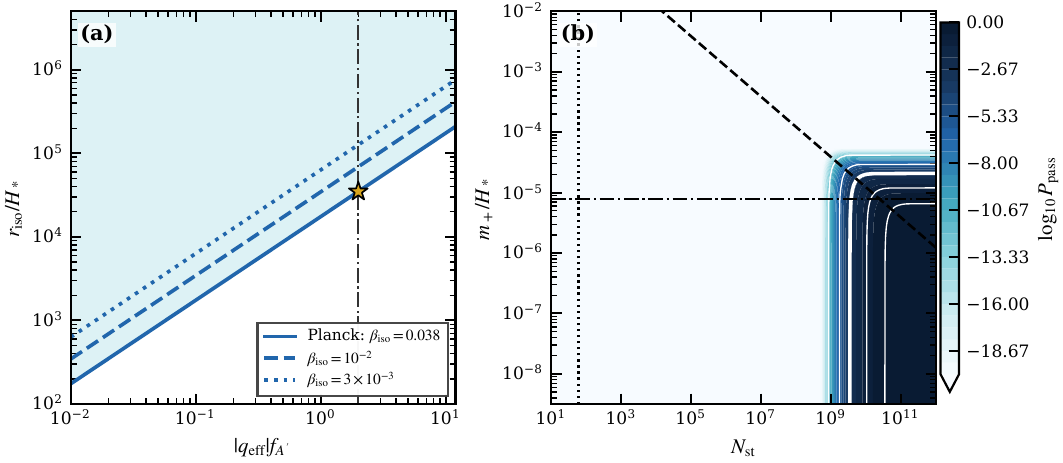}
\caption{CMB response and stochastic duration. The left panel gives the local displacement required by current and representative future uncorrelated-isocurvature sensitivities. The right panel gives the finite-duration pass probability for the all-DM quadratic conserved-inheritance response $|\qeff|f_{\Ap}=2$. The thick white contour marks $P_{\rm pass}=10^{-3}$. The $N_{\rm st}\sim60$ benchmark requires a coherent mean displacement or a response-reducing transition.}
\label{fig:entropy}
\end{figure*}

Once $N_{\rm st}$ and $H_*$ are specified with the inflationary radial potential the radial distribution gives the large-deviation weight. For $N_{\rm st}=60$ the CMB threshold is $h_{\rm iso}/\sigma_h\simeq2.8\times10^4$. Figure~\ref{fig:entropy} displays the probability surface.

\subsection{Scalar-sector consistency}
For the quartic branch we adopt a stochastic equilibrium prior. In the radiative estimate we identify the inflationary quartic with the late-time Abelian-Higgs quartic. The equilibrium distribution then gives the quartic constraint below. For an inflationary potential $V=\lambda_+(\hh^2)^2/4$ with $N_h$ real dark-Higgs degrees of freedom the radial expectation value is

\begin{equation}
\langle r^2\rangle=H_*^2\left(\frac{3}{2\pi^2\lambda_+}\right)^{1/2}
\frac{\Gamma[(N_h+2)/4]}{\Gamma[N_h/4]} .
\end{equation}

For the case of a complex dark Higgs, the corresponding typical-equilibrium estimate implies the quartic bound

\begin{equation}
\lambda_+<\frac{5.19\times10^{-19}}{(\qeff f_{\Ap})^4}.
\label{eq:lambdabound}
\end{equation}

For the perturbative all-dark-matter branch, the quartic equilibrium prior gives \(\lambda_+<3.24\times10^{-20}\), which is the required inflationary flatness scale of the radial direction.

Identifying the inflationary quartic with the late-time Abelian-Higgs quartic links this CMB scale to radiative stability in the minimal hidden sector. Gauge-loop corrections generate a quartic contribution of the approximate form

\begin{equation}
\Delta\lambda_h\simeq \frac{C_g g_D^4}{16\pi^2},
\label{eq:radiative}
\end{equation}

where $C_g$ denotes the gauge-sector loop coefficient. Requiring radiative stability then yields the upper bound

\begin{equation}
g_D\lesssim\frac{7.2\times10^{-5}}{|\qeff| f_{\Ap}}
\left(\frac{3}{C_g}\right)^{1/4}.
\label{eq:gdbound}
\end{equation}

For the benchmark perturbative case \(|\qeff| f_{\Ap}=2\) with \(C_g=3\), radiative stability gives \(g_D\lesssim3.6\times10^{-5}\). Ultraviolet symmetries or additional particle content can protect the quartic and enlarge this consistency window.

The decay epoch gives the complementary lower scale for the gauge coupling. In the relativistic-width limit \(x=m_{\Ap}^2/m_h^2\ll1\), the decay width satisfies \(\Gamma\simeq g_D^2m_h^3/(32\pi m_{\Ap}^2)\), leading to an independent lower bound on the gauge coupling once the required decay epoch is specified.

\subsection{Portal and Higgsed-vector consistency}
Low-energy portal limits and the minimal Higgsed completion provide the remaining realization conditions. Existing laboratory experiments, astrophysical observations, and stellar-cooling bounds constrain the familiar dark-photon parameter space characterized by \((m_{\Ap},\epsilon)\) \cite{Pospelov:2008jk,An:2013yfc,An:2013yua,Caputo:2021eaa}. Similarly, direct-detection searches and future sub-GeV experimental programs organize this parameter space according to experimental sensitivity \cite{Essig:2011nj,Essig:2022dfa,McDermott:2017qcg}. The scalar inheritance parameter \(\qeff\) adds an independent origin coordinate to this visible-sector parameter plane.

Visible-sector phenomenology and the scalar-response constraint define complementary coordinates. The AxionLimits repository and its associated dark-photon compilation summarize the experimental \((m_{A'},\epsilon)\) plane \cite{OHare:AxionLimits,Caputo:2021eaa}. The hidden gauge coupling \(g_D\) and the visible kinetic mixing \(\epsilon\) are displayed on separate axes because any relation between them requires an explicit ultraviolet matching prescription. A laboratory discovery would determine low-energy parameters, while the calculated response map adds information about the production history.

The Higgs origin of the vector mass introduces an additional theoretical consistency requirement. For a renormalizable Higgs potential, sufficiently early hidden-sector symmetry breaking requires \cite{Kitajima:2024vbc}

\begin{equation}
\frac{m_{\Ap}}{g_D}\gg60\,\mathrm{eV}
\left(\frac{2\pi}{\lambda_h}\right)^{1/4} .
\label{eq:Kitajima}
\end{equation}

Substituting the CMB-motivated quartic value \(\lambda_h=3.24\times10^{-20}\) gives the numerical scale \(7.1\,\mathrm{MeV}\) on the right-hand side. For sub-eV Higgsed vectors, this scalar sector requires a correspondingly small hidden gauge coupling when it governs both inflationary flatness and late-time spontaneous symmetry breaking.

The early-symmetry-breaking condition and the radiative stability criterion are independent theoretical conditions. Together they determine whether an abundance-matched parameter point admits a technically consistent Higgsed completion. Minimal Higgsed realizations occupy a joint consistency window. In this window the scalar quartic satisfies the CMB flatness condition. Radiative corrections scale as \(\Delta\lambda_h\sim g_D^4\) and the scalar decay ends at the chosen cosmological epoch.

The response map supports four cosmological realizations. A subdominant vector fraction scales the CMB response by $f_{\Ap}$. A coherent inflationary prior provides the required displacement. Transition dynamics can reduce $\partial\ln\Omega_{\Ap}/\partial\ln r$ before number freeze-out. The daughter momentum determines the nonrelativistic epoch.

Hidden-sector thermalization and number-changing reactions can drive the final yield toward an attractor with \(\qeff\to0\) \cite{Hall:2009bx,Bellomo:2022pav,Holst:2023lrm,Franciolini:2026pbhi}. Annihilation plateaus and entropy dilution can produce an analogous response. The abundance normalization and its logarithmic derivative then provide two independent model-building targets.

\section{Conclusions}
The Higgs origin of dark-photon dark matter appears in the derivative structure of its abundance map. For a smooth local map the separate-universe relation converts the radial fluctuation into CDM isocurvature. In the decoupled spectator regime of Eq.~\eqref{eq:spectatorconditions} this relation gives the uncorrelated spectrum in Eq.~\eqref{eq:master}. The finite inflaton--dark-Higgs mixing enters through the cross-spectrum in Eq.~\eqref{eq:crossspectrum}. Within quadratic conserved-number inheritance, displacement-independent transfer gives $q_{\rm eff}=2$. A nondecreasing transfer factor gives $q_{\rm eff}\ge2$. The crossover calculation evaluates the remaining transfer Jacobian. It contains positive, negative, and near-zero responses at finite transferred abundance. Hyperbolic-tangent, error-function, and thermal-mass transitions retain all three branches. Finite inflationary quartics shift their locations continuously.

The calculated $q_{\rm eff}$ maps the uncorrelated Planck limit onto a surface for $v_h/H_*$. The response-reduction corridor lowers the required symmetry-breaking scale by more than two orders of magnitude relative to the amplified benchmarks while retaining finite comoving action. Stochastic duration, radiative stability, early symmetry breaking, and daughter-particle coldness then define realization regions on this surface.

Low-energy measurements determine $(m_{A'},\epsilon)$, while CMB isocurvature constrains the dependence of the relic abundance on the dark-Higgs initial condition and transfer history. The two observables can differentiate Higgsed-vector cosmologies with identical present-day particle properties.

\section*{Acknowledgements}
SC acknowledges the Istituto Nazionale di Fisica Nucleare (INFN) Sez. di Napoli,  Iniziative Specifiche QGSKY and MoonLight-2  and the Istituto Nazionale di Alta Matematica (INdAM), gruppo GNFM, for the support. This paper is based upon work from COST Action CA21136 -- Addressing observational tensions in cosmology with systematics and fundamental physics (CosmoVerse), supported by COST (European Cooperation in Science and Technology). This research was funded by the National Natural Science Foundation of China (NSFC) under Grant No. U2541210 and 12505238.

\appendix

\section{Response-flow and action derivation}
\label{app:responseflow}

For the radiation-dominated benchmark, $\mathcal H=H/m_h$ and $\mathcal H'/\mathcal H=-2$. Writing $z=x'$, the background and tangent systems are
\begin{align}
x'&=z, & z'&=-z-\frac{U_{,x}}{\mathcal H^2},\\
u_x'&=u_z, & u_z'&=-u_z-\frac{U_{,xx}}{\mathcal H^2}u_x.
\end{align}
The initial tangent data follow from $x(N_i)=r_i/v_h$ and the light-field slow-roll velocity. Once the transition potential is stationary, the orbit energy and its response are
\begin{align}
E&=\frac12\mathcal H^2z^2+U_{\rm br}(x),\\
\frac{\partial E}{\partial\ln r_i}&=\mathcal H^2zu_z+U_{{\rm br},x}u_x.
\end{align}
For a periodic one-dimensional orbit, $J_{,E}=1/\omega$. Conservation of $a^3J$ in the adiabatic regime and field-independent perturbative decay yield Eq.~\eqref{eq:qdynamic}.

\begin{figure*}[!t]
\centering
\includegraphics[width=\textwidth]{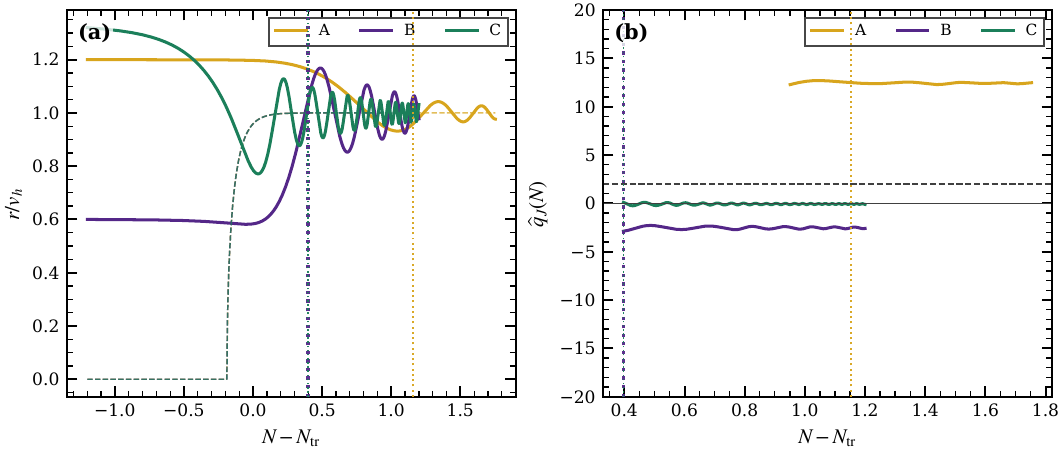}
\caption{Time-resolved transition histories for Table~\ref{tab:dynamic}. Panel (a) gives the radial trajectories. Dashed curves show the instantaneous positive minimum. Panel (b) gives the action-response diagnostic $\widehat q_J(N)$. Colored dotted lines mark the adiabatic handoff at $H/\omega=0.1$. The late-time plateaus give $q_{\rm eff}$.}
\label{fig:responseflow}
\end{figure*}

During the transition we display the diagnostic
\begin{equation}
\widehat q_J(N)\equiv\frac{\partial\ln J[E(N)]}{\partial\ln r_i},
\end{equation}
once the broken-phase action is defined. Its asymptotic value is $q_{\rm eff}$. Figure~\ref{fig:responseflow} shows this evolution. Colored dotted lines mark the first time each trajectory reaches $H/\omega=0.1$.

\section{Transition-profile and quartic dependence}
\label{app:robustness}

The central map uses the tanh profile in Eq.~\eqref{eq:smoothtransition}. We evaluate two independent deformations. The first replaces $s(N)$ by an error function. The second uses the radiation-era thermal potential
\begin{equation}
U_T(x,N)=\frac{(x^2-1)^2}{8}+\frac14 e^{-2N}x^2,
\label{eq:thermalprofile}
\end{equation}
whose curvature at the origin changes sign at $N=0$, following $m_T^2\propto T^2\propto a^{-2}$ \cite{Quiros:1999jp}. At a finite thermal handoff, the potential has a minimum at $x_{\min}=\sqrt{\alpha}$ with $\alpha=1-e^{-2N}$. Writing $E_{\rm rel}$ for the energy above the moving minimum and rescaling $x=\sqrt{\alpha}\,y$ gives $J_T=\alpha^{3/2}J_{\rm br}(E_{\rm rel}/\alpha^2)$, which transports the orbit adiabatically to the zero-temperature action. The second deformation retains Eq.~\eqref{eq:smoothtransition} and varies $\Lambda_+=\lambda_+v_h^2/m_h^2$.

\begin{figure*}[!t]
\centering
\includegraphics[width=\textwidth]{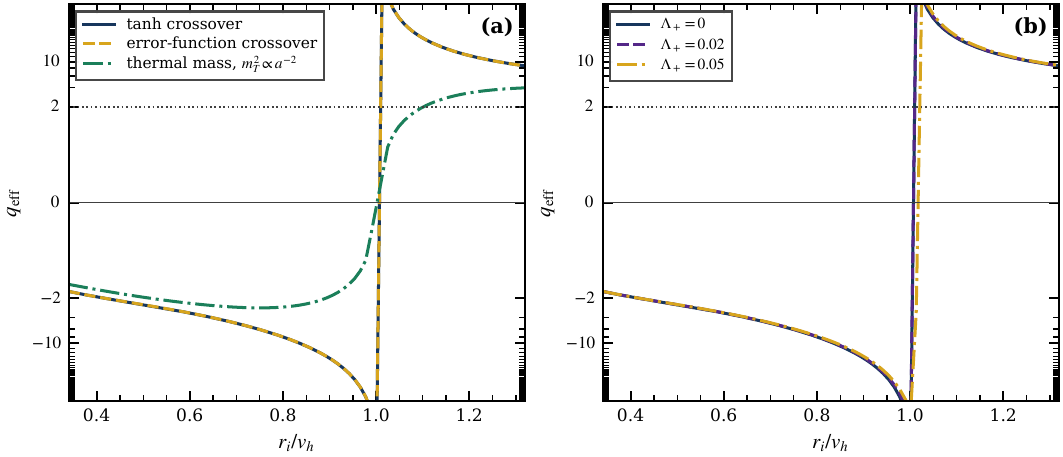}
\caption{Dynamical-response dependence at $m_h/H_{\rm tr}=1$. (a) Hyperbolic-tangent, error-function, and thermal-mass transitions. (b) Inflationary-quartic deformations of the tanh crossover. The horizontal lines mark $q_{\rm eff}=0$ and the quadratic inheritance value $q_{\rm eff}=2$.}
\label{fig:robustness}
\end{figure*}

Figure~\ref{fig:robustness} compares the three transition profiles and the quartic deformations at $m_h/H_{\rm tr}=1$. Every profile contains positive, negative, and zero-response branches. The tanh and error-function curves nearly coincide. The thermal mass shifts the response scale while retaining the sign-reversal topology. Values $\Lambda_+=0.02$--$0.05$ move the transition locus continuously. Tangent responses for all four deformation classes agree with independent five-point derivatives to relative precision better than $8.1\times10^{-9}$.

\section{Numerical validation}
\label{app:numerics}
The post-transition action uses a 1400-point quadrature table and a shape-preserving cubic interpolant. At $E=10^{-8}$, the ratio $J/E=1.0000000075$ reproduces the harmonic limit. The response surface contains 1596 controlled trajectories satisfying the single-basin and adiabatic-handoff criteria stated in Sec.~\ref{sec:dynamical}. Eight validation points spanning amplified, sign-reversed, and response-reduced regimes were recomputed with symmetric five-point derivatives in $\ln r_i$ at four step sizes. The maximum relative tangent--finite-difference discrepancy is below $2.0\times10^{-8}$. The transition-profile and quartic deformations provide an independent four-model validation set with maximum relative discrepancy below $8.1\times10^{-9}$.

\begin{figure}[!t]
\centering
\includegraphics[width=\columnwidth]{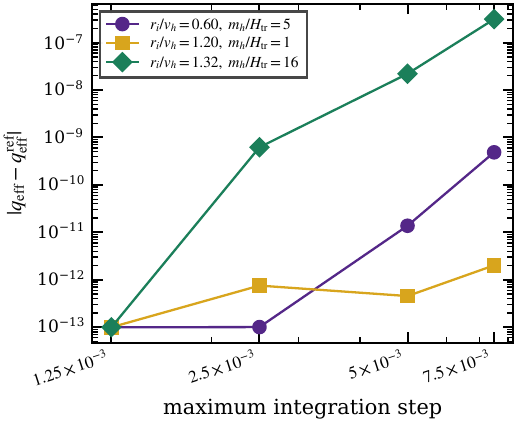}
\caption{Response convergence as the maximum integration step is reduced. The reference values use the tightest tolerance and smallest integration step.}
\label{fig:convergence}
\end{figure}

Figure~\ref{fig:convergence} shows integration-step convergence for the representative trajectories in Table~\ref{tab:dynamic}. These checks give stable response values across the amplified, sign-reversed, and response-reduced branches.

\bibliographystyle{apsrev4-2}
\bibliography{refs1}

\end{document}